\documentclass[%
 reprint,
 amsmath,amssymb,
 aps,
 pra,
]{revtex4-1}

\usepackage{graphicx}% Include figure files
\usepackage{dcolumn}% Align table columns on decimal point
\usepackage{bm}% bold math
\usepackage{natbib}
\usepackage{hyperref}% add hypertext capabilities
\usepackage{siunitx}
\usepackage[caption=false]{subfig}
\usepackage{booktabs}
\usepackage[export]{adjustbox}
\usepackage{multirow}
\usepackage{physics}
\usepackage{accents}
\begin{document}

%\preprint{APS/123-QED}

\title{Scalable quantum computation with fast gates in two-dimensional microtrap arrays of trapped ions}% Force line breaks with \\

\author{Zain Mehdi}
 \email{zain.mehdi@anu.edu.au}%
 \affiliation{Department of Quantum Science, Research School of Physics, Australian National University}%
\author{Alexander K. Ratcliffe}
 \affiliation{Department of Quantum Science, Research School of Physics, Australian National University}%
\author{Joseph J. Hope}
 \affiliation{Department of Quantum Science, Research School of Physics, Australian National University}%

\date{\today}% It is always \today, today,

\begin{abstract}
We theoretically investigate the use of fast pulsed two-qubit gates for trapped ion quantum computing in a two-dimensional microtrap architecture. In one dimension, such fast gates are optimal when employed between nearest neighbours, and we examine the generalisation to a two-dimensional geometry. We demonstrate that fast pulsed gates are capable of implementing high-fidelity entangling operations between ions in neighbouring traps faster than the trapping period, with experimentally demonstrated laser repetition rates. Notably, we find that without increasing the gate duration, high-fidelity gates are achievable even in large arrays with hundreds of ions. To demonstrate the usefulness of this proposal, we investigate the application of these gates to the digital simulation of a $40$-mode Fermi-Hubbard model.  This also demonstrates why shorter chains of gates required to connect arbitrary pairs of ions makes this geometry well suited for large-scale computation. 
\end{abstract}

% PACS, the Physics and Astronomy Classification Scheme.

\maketitle

%\tableofcontents

\section{\label{sec:introduction}Introduction}
Trapped ion platforms are very promising for implementing large-scale quantum computations, such as simulations of many-body quantum systems, in the near future. The achievement of a scalable quantum computer would allow for unprecedented advances in quantum chemistry, physics, and biology; from the study of molecular bonds and structure to better understanding high-temperature superconductivity. \par
While trapped ion platforms have demonstrated several key elements required for large-scale quantum computing - long coherence times \cite{Monroe2013}, single-qubit and two-qubit gates with fidelities above fault-tolerant thresholds \cite{Harty2014, Gaebler2016, Ballance2016}, and high-fidelity readout \cite{Myerson2008} - scalability remains an issue. Chains of ions are typically trapped in a common potential generated by a Paul trap, with collective vibrational modes used for entanglement. However, current approaches to scaling these devices are limited, either by speed or fidelity. This is largely due to the sideband-resolving mechanisms generally used to implement two-qubit entangling gates, that require the gate be much slower than the trap frequency, such as the Cirac-Zoller \cite{Cirac1995, Jonathan2000, Schmidt-Kaler2003b, Schmidt-Kaler2003c} and M\o{}lmer-S\o{}rensen \cite{Molmer1999, Sorensen1999, Sorensen2000} schemes. Moreover, as the number of ions in the chain is increased, the motional sidebands become harder to address, and thus the gate time must increase \cite{Zhu2006}.  \par

One proposal for scaling these devices involves two-dimensional arrays where ions are shuttled around segmented traps to perform multi-qubit gates \cite{Kielpinski2002,Amini2010,Kaufmann2014}. However these approaches require time-dependent manipulation of the trapping potentials, and even the fastest experimentally demonstrated processes take multiple trap periods. Recent work by Ratcliffe \textit{et al.}~\cite{Ratcliffe2018a} has proposed the use of fast entangling gates that can be performed between ions in neighbouring microtraps in a linear array using sequences of ultra-fast counter-propagating pairs of laser pulses \cite{Garcia-Ripoll2003,Duan2004a,Bentley2013,Bentley2015b,Ratcliffe2018a}, even in the presence of large numbers of surrounding ions. Notably this scheme requires no manipulation of trapping potentials for shuttling ions; entangling gates can be performed between ions in neighbouring traps \textit{in situ} and outperform state-of-the-art shuttling schemes in terms of speed. Further work has demonstrated that these gates can be enhanced by the presence of micromotion, which is otherwise detrimental to gates implemented on radial trap modes \cite{Ratcliffe2020}. \par

\begin{figure}[t!]
    \centering
    \includegraphics[width=0.8\columnwidth]{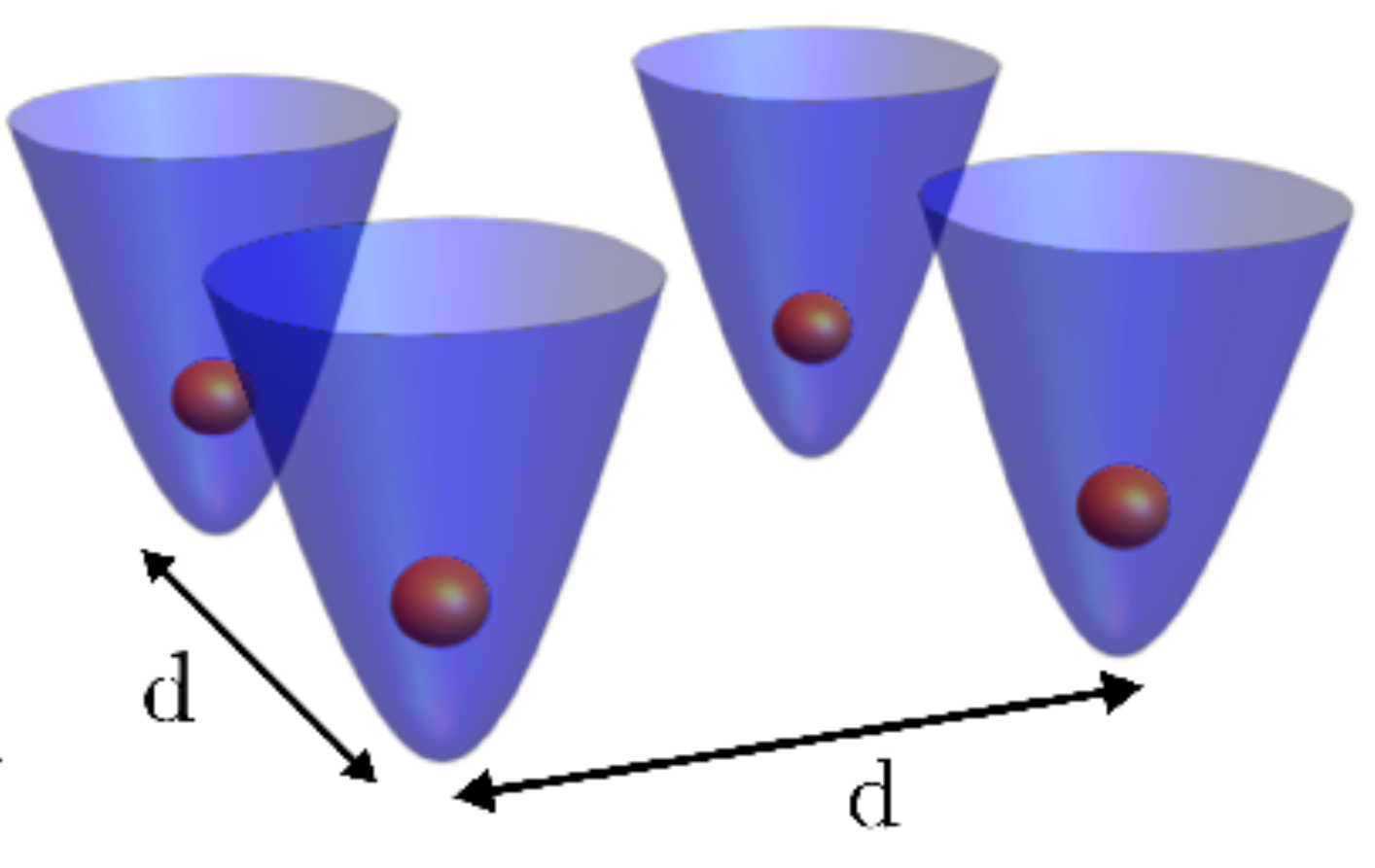}
    \caption{Schematic of a simple $2\times 2$ ion crystal in two-dimensional. Ions are represented as (orange) spheres in (blue) potentials. The distance between the minima of nearest-neighbour traps, $d$, is shown. In this manuscript, we assume the microtraps to be Paul traps with equal radial and axial trapping frequencies of $\omega_t$, each containing a single ion.}
    \label{fig:2x2_schematics}
\end{figure}

These recent developments lay the basis for this work, where we investigate the application of fast entangling gates to architectures where ions are individually trapped in a two-dimensional microtrap array \cite{Kumph2016,Hakelberg2019}. A simple $2\times 2$ microtrap array is visualised in Fig.~\ref{fig:2x2_schematics}. In Section \ref{sec:background}, we introduce the mechanism considered for implementing fast entangling gates that uses optimised sequences of ultra-fast $\pi$ pulses. In Section \ref{sec:microtrap_array} we show that high-fidelity gate operations are theoretically achievable faster than the trap period, for realistic trap parameters and demonstrated laser repetition rates. These gates are insensitive to the number of surrounding ions, paving the way for computation in large microtrap arrays. As an example of a large-scale computation that fast gates in microtrap arrays could enable, we study the feasibility of realising a digital simulation of the Fermi-Hubbard model in Section \ref{sec:simulation}.

\section{\label{sec:background}Background: Fast entangling gates with ultrafast pulses}
Entangling gates can be implemented in trapped ions faster than the trapping period $2\pi/\omega_t$ by exciting multiple collective motional modes, and using state-dependent motion to generate entanglement between their electronic states. This has recently been demonstrated by Sch\"{a}fer \textit{et al}.~\cite{Schafer2018} using amplitude-shaped pulses to drive state-dependent trajectories that implement maximally entangling gates \cite{Steane2014}. While this method was able to implement a high-fidelity ($99.8\%$) entangling gate in $1.6$ microseconds, gates significantly faster than the trap frequency were associated with a much lower fidelity (${\sim}60\%$).\par
An alternative mechanism for fast entangling two-qubit gates has the state-dependent trajectories driven by counter-propagating pairs of ultra-fast resonant pulses each with pulse area $\pi$ that implement geometric phase gates \cite{Garcia-Ripoll2003, Duan2004a, Bentley2013, Bentley2015b, Bentley2016, Ratcliffe2018a, Hussain2016, Heinrich2019}. Each pulse pair imparts a state-dependent momentum kick of $\pm 2\hbar k$ on each ion. When interspersed with periods of free evolution, the state-dependent kicks can be used to orchestrate internal state-dependent phase-space trajectories of multiple motional modes. In the ideal two-qubit fast gate, these phase-space loops will close perfectly and the state-dependent phase accumulation along each trajectory will lead to a $\pi/2$ phase difference between the $\ket{\uparrow,\uparrow}$/$\ket{\downarrow,\downarrow}$ and $\ket{\uparrow,\downarrow}$/$\ket{\uparrow,\downarrow}$ internal basis states. However, designing pulse sequences to orchestrate these trajectories such that they satisfy these two conditions is highly-nontrivial and requires a numerical optimisation approach. \par

In this manuscript, we consider fast gate schemes where pulse sequences have pulse pairs grouped together, with pulse groups separated by periods of free evolution. We employ a global procedure to optimise pulse sequences for high-fidelity fast gates \cite{Gale2020}. In this optimisation the free parameters are the number of pulses in each group (corresponding to the magnitudes of the state-dependent kicks). The total gate time is fixed in each optimisation, characterises the duration of free evolution between pulse groups. We will report the results of these optimisations in terms of the \textit{minimum resolving repetition rate} $f_\text{min}$, which is the minimum repetition rate required such that pulse groups do not overlap. It has previously been shown that for a laser with a pulse rate above $f_\text{min}$, the gate fidelity is robust to the finite repetition period between $\pi$ pulses \cite{Ratcliffe2018a}. Pulse sequences can be optimised for a specific repetition rate with a simple extension of our optimisation procedure, as outlined in Ref.~\cite{Gale2020}. Further detail of this procedure is provided in Appendix \ref{append:optmethods}.

\section{Results: Fast gates in a 2D microtrap array \label{sec:microtrap_array}}
\begin{figure}[t!]
    \centering
    \includegraphics[width=0.5\textwidth]{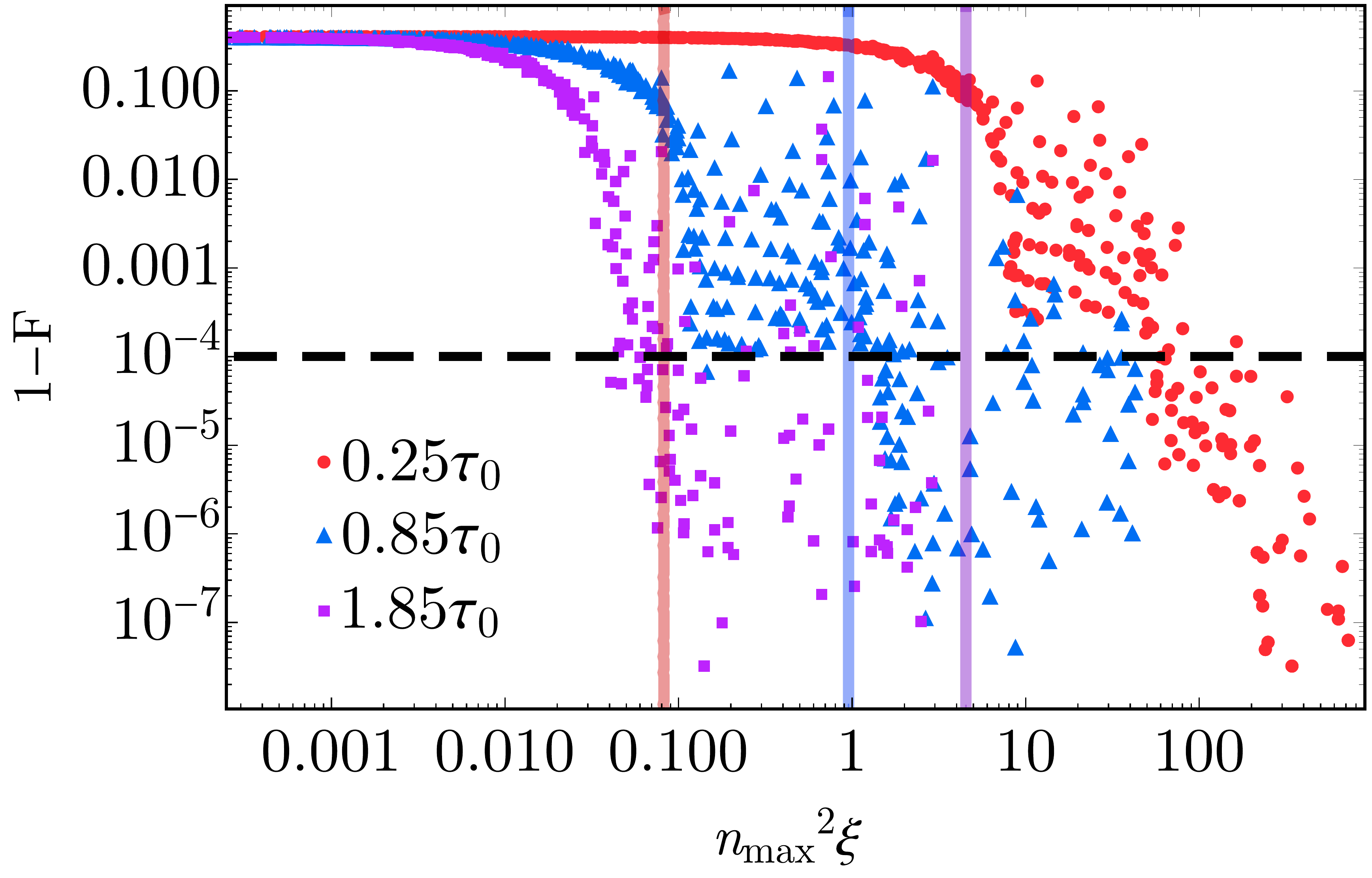}
    \caption{Infidelities of optimised gates are plotted as a function of $n_{max}^2\xi$, where $n_{max}$ is the largest number of pulse pairs in a given group for a given gate, plotted for several different gate times (in multiples of the trap period $\tau_0 = 2\pi/\omega_t$).  For each gate time, the infidelity of a gate is well described by the parameter $n_{max}^2\xi$ until it falls below approximately $1-\text{F} = 10^{-2}$.  The vertical lines correspond to the maximum values of $n_{max}^2\xi$ achievable with a state-of-the-art $5$~GHz repetition rate laser \cite{Heinrich2019} without pulse groups overlapping for a given gate time and a trap geometry with $\omega_t = 2\pi \times 1.2$ MHz and $d=100$ microns ($\xi = 1.2\times10^{-4})$. The horizontal dashed line is an indicative fidelity threshold to implement fault-tolerant error correction with a Bacon-Shor code with a depth of $10$ \cite{Cross07_ErrorCorr}.}
    \label{fig:nsqxivsinf}
\end{figure}
\begin{figure}[t!]
    \centering
    \includegraphics[width=0.46\textwidth]{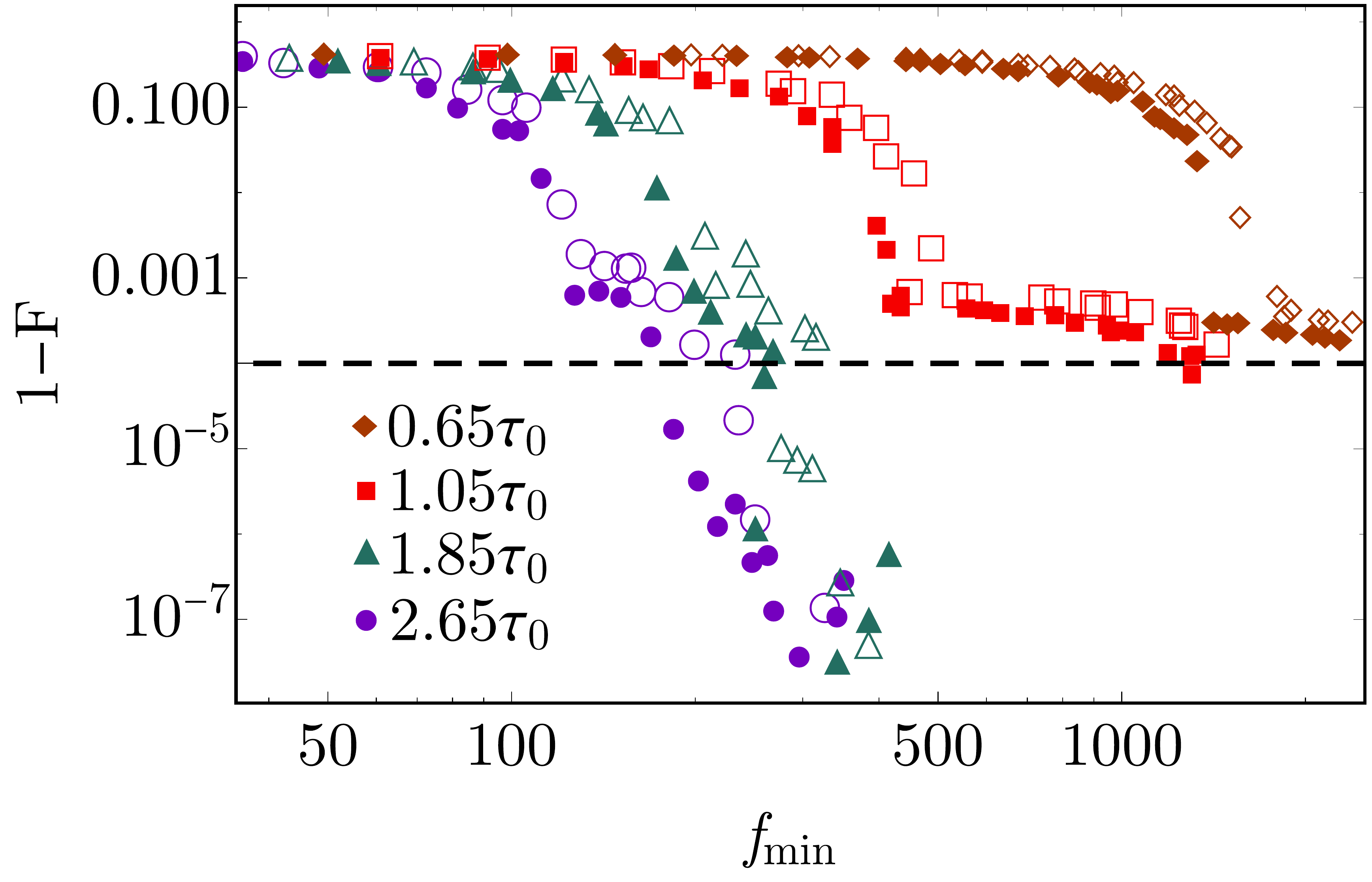}
    \caption{Gate infidelity as a function of resolving repetition rate for $\xi=1.2\times10^{-4}$; which corresponds to an inter-trap distance of $d=100\,\mu$m for $\omega_t=2\pi\times1.2\,$MHz. Results are presented for a variety of gate times, for gates between nearest-neighbour (filled) and diagonal (empty) pairs of ions in a $2\times2$ cell. Both repetition rate and gate times are presented in trap periods ($\tau_0=2\pi/\omega_t$).  The target error correction threshold $1-F=10^{-4}$ is shown by the horizontal dashed line.}
    \label{fig:2Dsym_reprate}
\end{figure}

We first examine a simple $2\times2$ square cell of ions in individual traps with an edge length $d$ corresponding to the distance between nearest-neighbour traps, as shown in Figure \ref{fig:2x2_schematics}. As there is only one key timescale of this system, the angular trap frequency $\omega_t$, the results of this section will be presented in units of $\frac{\omega_t}{2\pi}$ (trap frequencies) and $\frac{2\pi}{\omega_t}$ (trap periods). In the analogous 1D case, the mode structure is defined by the scaled difference between the breathing and common motional modes frequencies, $\chi \equiv \frac{\omega_{BR}-\omega_t}{\omega_t}$ which can be calculated in terms of trap parameters. Equivalently, we find that this 2D cell can be non-dimensionalised in terms of the normalised difference between the \textit{squared} breathing and common motional mode frequencies $\xi\equiv\frac{\omega_{BR}^2-\omega_t^2}{\omega_t^2}$ (see Supplemental Material to Ref.~\cite{Ratcliffe2018a}). This can similarly be calculated based on trap parameters, as outlined in Appendix \ref{appendix:non-dimensionalisation}. 
 \par
We perform global optimisations of pulse sequences to realise a fast gate between nearest-neighbour ions in the $2\times 2$ square cell for different values of $\xi$. In analogy to the 1D treatment in Ref.~\cite{Ratcliffe2018a}, we identify $n_{max}^2\xi$ as a characteristic parameter to characterise gate dynamics, where $n_\text{max}$ is the number of pulses in the largest pulse group of a given optimised pulse sequence. In our calculations we assume the common-motional Lamb-Dicke parameter is fixed to $\eta = \sqrt{\frac{\hbar}{2M\omega_t}} = 0.16$; this corresponds to a fixed trap frequency of $\omega_t/2\pi = 1.2$~MHz if the $\pi$-pulses are performed on the $S_{1/2}\rightarrow P_{3/2}$ optical transition in $^{40}\text{Ca}^{+}$. Fig.~\ref{fig:nsqxivsinf} shows gate infidelity is monotonic with $n_{max}^2\xi$ until it falls below $10^{-2}$.  In the monotonic region, the infidelity is dominated by phase accumulation error, for which optimal solutions are well characterised by $n_{max}^2\xi$.  When this parameter is high enough to satisfy the phase condition, residual infidelity is due to motional restoration errors, which are not well characterised by $n_{max}^2\xi$.  \par
\begin{figure*}[t!]
    \centering
    \includegraphics[width=\textwidth]{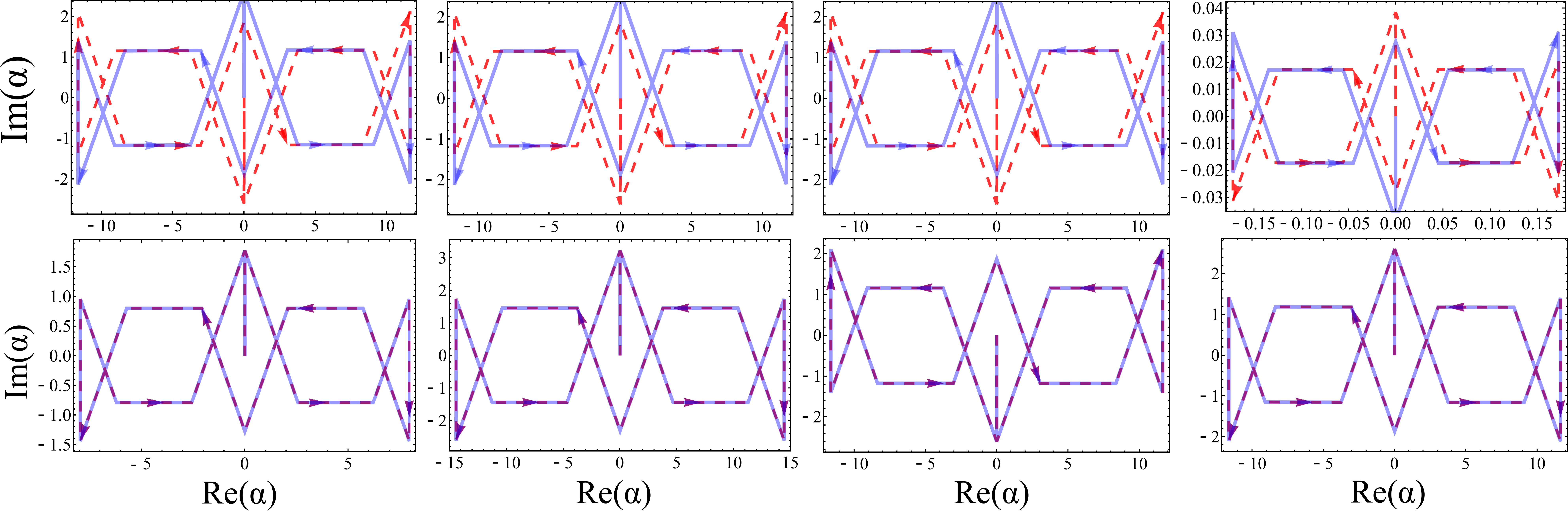}
    \caption{Phase-space trajectories of the motional modes of a fast gate in a $2\times 2$ microtrap array. The real and imaginary components of $\alpha$ correspond to non-dimensional position and momentum of each mode, respectively: $\alpha = x/(2x_0)+ip/(2p_0)$ where $x_0=\sqrt{\hbar/2M\omega_t}$ and $p_0=\sqrt{M\hbar\omega_t/2}$. Each motional modes is represented in the rotating frame with respect to its mode frequency. The red (dashed) lines correspond to the trajectories when both ions have the same qubit state ($\ket{\uparrow\uparrow}$,$\ket{\downarrow\downarrow}$), and the blue (solid) lines correspond to trajectories where the two targeted ions are in different qubit states ($\ket{\downarrow\uparrow}$,$\ket{\downarrow\uparrow}$). Note that each motional mode is restored by end of gate operation. The difference in areas enclosed by these two trajectories will lead to state-dependent phase accumulation; when the sum of the area-differences are $\pi/2$, the gate is maximally entangling.  In the absence of pulse errors, this particular gate operation has a fidelity of approximately $1-10^{-9}$ and a duration of $2.0$ trap periods, for a trap characterised by $\xi = 1.2\times10^{-4}$.  }
    \label{fig:PS_traj}
\end{figure*}

 
Our results suggest that for realistic trapping parameters of $\omega_t=2\pi\times1.2$~MHz and $d=100\,\mu$m (realistic values for the experimental `Folstrom' microtrap array reported in Ref.~\cite{Kumph2016}), a state-of-the-art $5$~GHz repetition rate laser \cite{Heinrich2019} is able to resolve pulse sequences that implement gates as fast as $700$~ns ($0.85$ trap periods) with infidelities as low as ${\sim}10^{-4}$. This is in agreement with Fig.~\ref{fig:2Dsym_reprate}, which shows optimisations of gates as fast as $0.65$ trap periods able to achieve almost $99.99\%$ fidelity, requiring repetition rates ${\sim}~2$~GHz. Longer gate times above $1-2$ trap periods require significantly lower repetition rates to achieve high-fidelities; Fig.~\ref{fig:2Dsym_reprate} demonstrates that a repetition rate of ${\sim}300$~MHz is sufficient to resolve $1.85$ trap period gates with fidelities above $99.99\%$, both for gates between nearest-neighbour ions, and ions in diagonally separated microtraps (i.e. with inter-trap distance of $\sqrt{2}d$). In general, low repetition rate lasers can be used for gates in this architecture, at the cost of longer gate times (as we show in the following sections, see Fig.~\ref{fig:diagvsadj_micro}). We emphasise that for smaller inter-trap distances $d$, these gate speeds can be achieved with lower repetition rate lasers. The phase-space trajectories of the motional modes during an exemplary $2.0$ trap period gate are visualised in Figure \ref{fig:PS_traj}. \par
While Fig.~\ref{fig:2Dsym_reprate} shows that gate fidelity is roughly monotonic with repetition rate, we see a discontinuity in the trends at roughly $1-F{\sim}10^{-3}$, particularly apparent for shorter gate times. Like the discontinuity observed in Fig.~\ref{fig:nsqxivsinf}, this happens as the infidelity transitions from being limited by the accumulation of sufficient phase difference between the internal states (low repetition rate), and motional restoration (high repetition rate).  The threshold depends on the gate time and $\xi$, but when all else is equal, we see that a higher repetition rate is always better. For sub-trap-period gate times, we see that motional restoration errors contribute to the infidelity on the order of $10^{-3}$.  \par
We have become aware of recent work by Wu and Duan \cite{wu2020twodimensional}, where the authors investigate fast gates in a similar 2D microtrap architecture. Our results are comparable to those presented in their work; for system parameters corresponding to $\xi=1.2\times10^{-3}$ ($d=50~\mu$m,  $\omega_t/2\pi=0.93~$MHz - see Supplementary Material V. of \cite{wu2020twodimensional}) the authors describe a $2.15~\mu$s gate with infidelity ${\sim}10^{-4}$ with a $80~$MHz repetition rate using $86$ $\pi$-pulse pairs. For the same parameters, we find a comparable $1.93~\mu$s gate with an infidelity of $3.5\times10^{-4}$, requiring $f_{min}=88~$MHz and $132$ pulse pairs. We note that while the gate scheme used in Ref.~\cite{wu2020twodimensional} is optimal in creating the desired phase difference between internal states, it satisfies motional restoration by requiring gate times that are integer multiples of two trap periods. In contrast, our scheme allows arbitrary gate times, including those below the trap period, as we explicitly optimise both for desired phase and motional restoration of all modes.  Achieving both of these conditions for sub-trap period gates comes at the cost of requiring more pulses and higher laser repetition rate. The optimisation scheme applied in this manuscript places strict constraints on the timings of the pulses; results can be further improved by adding a second stage of local optimisations on pulse timings. We have reported on the effectiveness of such a two-stage optimisation protocols in Refs.~\cite{Gale2020,Mehdi2020Chain}.

\subsection{Performance in large microtrap arrays}
\begin{figure}[t!]
    \centering
    \includegraphics[width=0.45\textwidth]{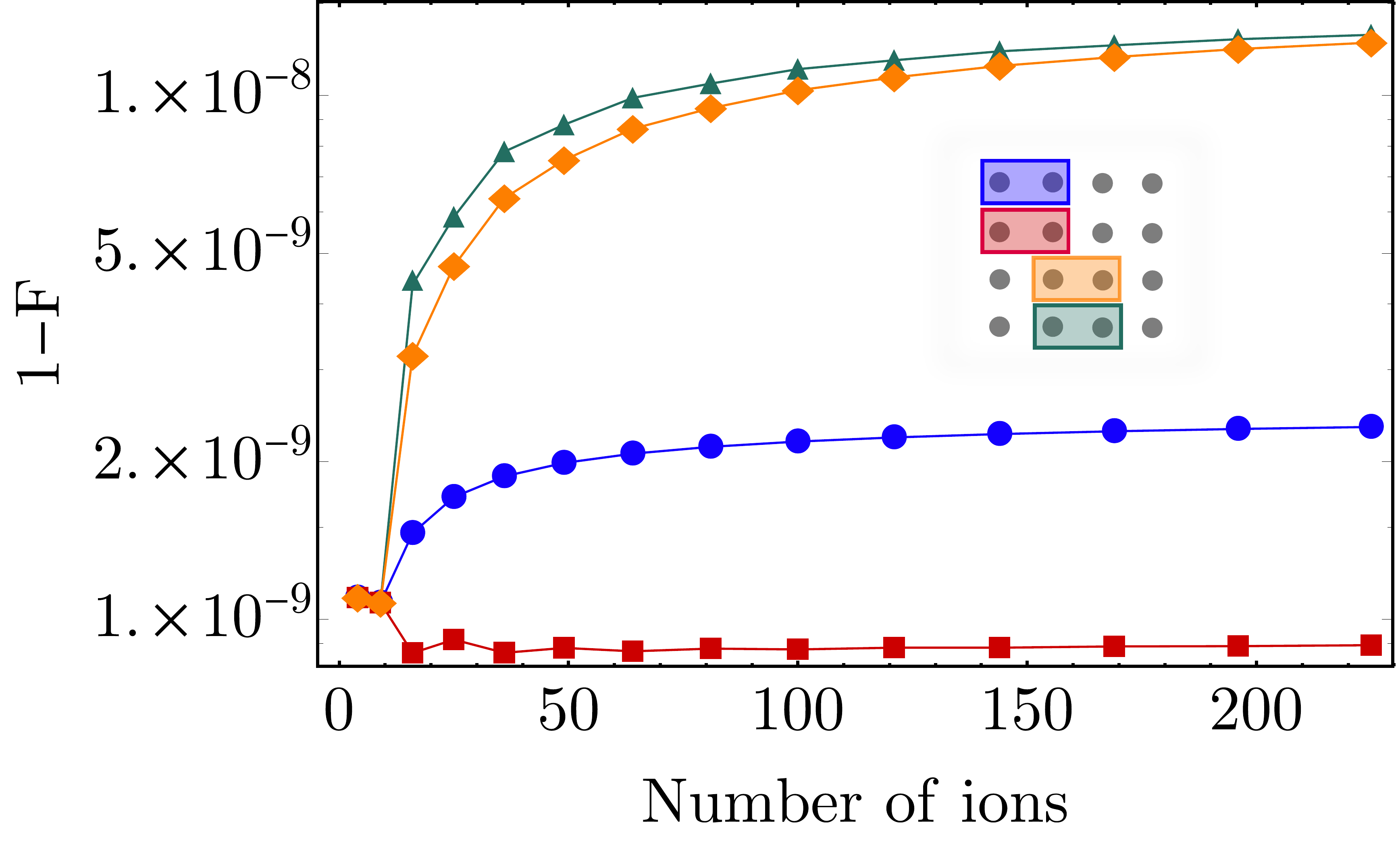}
    \caption{Scaling of nearest-neighbour fast gates in different locations in a square microtrap array with different numbers of ions. The pulse sequence has a total gate time of $2.0$ trap periods, a minimum resolving repetition rate of $450\frac{\omega_t}{2\pi}$, and a theoretical infidelity of $10^{-9}$ in a simple 2x2 array. The locations are pictorially represented for the $4\times 4$ array. Gate infidelity does not significantly increase, with minimal change from the infidelity of gates in a $2\times 2$ cell at best, and an order of magnitude increase (${\sim}10^{-8}$) at worst. }
    \label{fig:MicrotrapScaling}
\end{figure}
Thus far we have restricted our analysis to the simple microtrap array with four ions in a $2\times2$ cell. Here we will consider the performance of fast gates in scaled microtrap arrays; we will report results for $N\times N$ arrays, which will place an upper bound on achievable gate fidelities in more general $N\times M$ arrays ($N\geq M$). A brute force approach to this analysis might entail optimising gates individually for arrays of different sizes, and compare achievable fidelities for comparable operation times and repetition rates. However, the complexity of the infidelity expression given in Eq.~\eqref{eqn:2d infidelity truncated} scales with the number of motional modes, which in turn scales as $N^2$ (i.e. with the number of ions in the array). Thus optimisation with this cost-function quickly becomes computationally infeasible for all but the smallest 2D microtrap arrays. \par

We take an alternative approach, where we directly apply gates optimised for the simple $2\times2$ cell to larger $N\times N$ arrays, in analogy to the one-dimensional treatment in Ref.~\cite{Ratcliffe2018a}. This approach has the benefit of the control scheme (i.e. the pulse sequence used) remaining constant as the number of ions is increased, and thus the total operation time and required repetition rate also does not change. Calculation of the phase accumulation and motional restoration terms in Eq.~\eqref{eqn:2d infidelity truncated} includes all motional modes of the ions, and thus the fidelities we report here are a lower-bound on what is achievable in large microtrap arrays.\par
The results of this approach is shown in Fig.~\ref{fig:MicrotrapScaling}, which shows that gate is largely unaffected by the presence of many surrounding ions, with infidelity plateauing to extremely low values. This is similar to the scaling results presented in \cite{wu2020twodimensional}. We find that the magnitude to which the gate fidelity is affected depends on the location of the two ions the gate is performed on in the lattice. We find that in the best cases (along the edge of the array) the size of the lattice has little affect on the infidelity, and in the worst-cases (toward the middle of the lattice) the infidelity growing by little more than an order of magnitude (to ${\sim}10^{-8}$). Given that our optimisation routine is able to find extremely low infidelity solutions, gates even in the worst-case locations will be very robust in large microtrap arrays. This shows that microtrap arrays are well suited to large-scale quantum computation; which we discuss further in Section \ref{sec:simulation}.

\subsection{Optimal diagonal gates}

\begin{figure}[t!]
    \centering
    \includegraphics[width=0.42\textwidth]{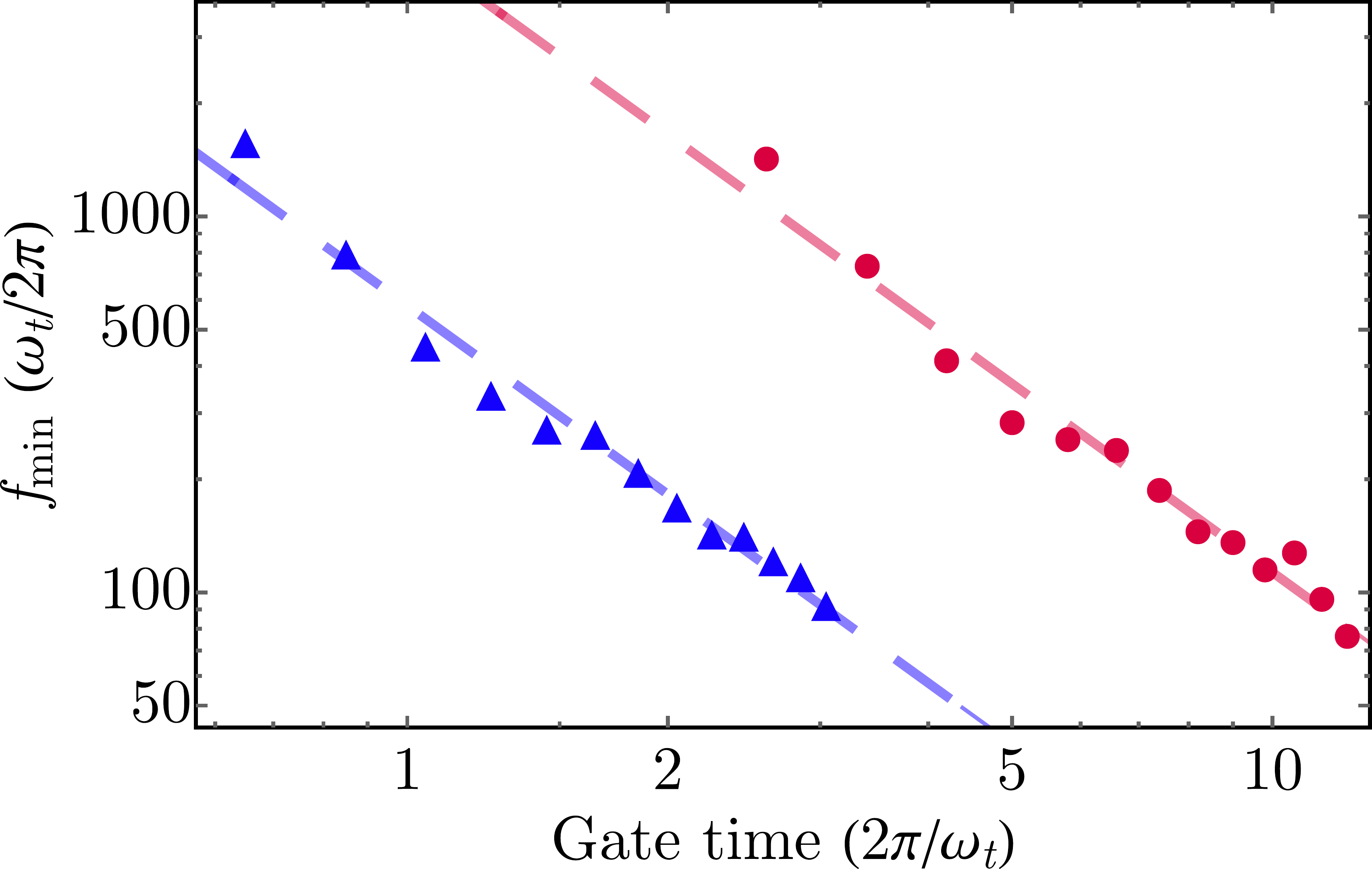}
\caption{Minimum repetition rates required to resolve gates with fidelity above $99\%$ as a function of gate time for a single diagonal gate (blue, triangles) and for an equivalent operation built out of nearest-neighbour gates (red, circles). Each data point corresponds to a gate optimised for a $2\times 2$ microtrap array with $\xi = 1.2\times10^{-4}$. Repetition rates and gate times are both presented in trap units; in multiples of $\frac{2\pi}{\omega_t}$ and $\frac{\omega_t}{2\pi}$, respectively.}
    \label{fig:diagvsadj_micro}
\end{figure}
Previous studies have shown that, for computation on a linear ion chain, it is optimal to perform fast gates exclusively between nearest-neighbour (NN) ion pairs \cite{Duan2004a,Bentley2015b}. This optimality is particularly important for gates faster than the trapping period, as gates between two distant ions must involve the motion of the ions between them, and thus are limited to a timescale set by the trapping frequency. Non-nearest neighbour operations can be built using SWAP operations which can be realised with three NN fast gates, up to local rotations \cite{Taylor2017}. However, it has not been shown whether NN optimality extends to gates between ions that are diagonally separated in the square $2\times2$ cell. The existence of an intuitive answer is obscured by the trade-off between increased gate time for (a) a single diagonal gate as compared to a NN gate due to a factor of $\sqrt{2}$ larger inter-ion separation, and (b) an equivalent NN operation due to the multiple gates required to build the SWAP operation(s). \par
We will seek to clarify this trade-off by comparing diagonal gates to their NN equivalent by the repetition rates required to resolve high-fidelity (above $99\%$) gates across a range of gate times. The fidelity and operation time for the equivalent NN operation is calculated by multiplying fidelities and adding the gate times of each of the four constituent NN operations (three gates to compose the SWAP operation, and one entangling gate). This is presented in Fig.~\ref{fig:diagvsadj_micro}, where we fit each data-set to the trend $f_\text{min}=a\tau_G^{-5/3}$ (where $f_\text{min}$ and $\tau_{G}$ are the laser repetition rate and gate time in trap units). The fitted parameter $a$ is $578.8$ and $5245.3$ for the diagonal gate and equivalent NN operation data, respectively. This suggests that the repetition rate requirement to perform a single diagonal gate is almost an order of  magnitude lower than the required repetition rate to implement an equivalent operation built out of NN gates, for a given operation time. Therefore it is clear that the notion of nearest-neighbour optimality does not extend to gates between diagonally separated ions in a microtrap arrays.

\section{Example computation: Simulation of the Fermi-Hubbard model \label{sec:simulation}}
\begin{figure*}[t!]
    \centering
    \includegraphics[width=\textwidth]{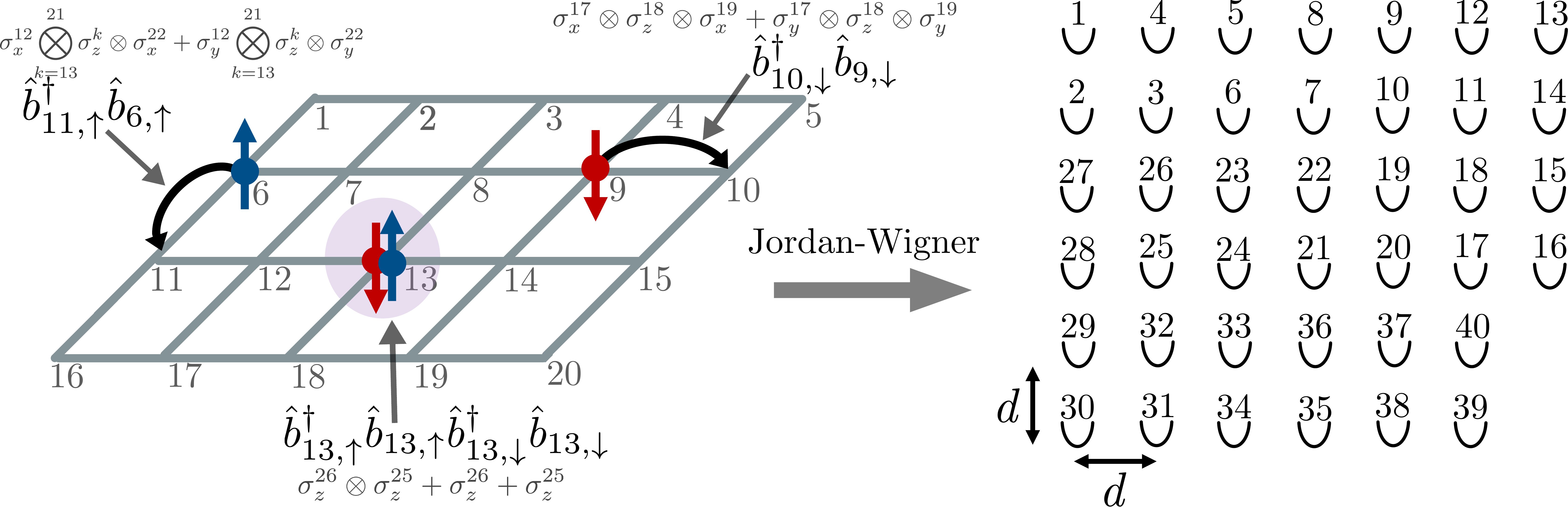}
    \caption{Mapping of a $20$-site Fermi-Hubbard model to a system of $40$ interacting ion-qubits in a 2D microtrap array. We have deliberately chosen the numbering of the qubits to the ions in the array to reduce the number of gates required to simulate Fermi-Hubbard dynamics. Exemplary terms in the Fermi-Hubbard Hamiltonian are shown representing different types of dynamics: hopping between nearest neighbour sites (black arrows), and onsite interaction (purple) between spin up and spin down fermions. The Jordan-Wigner (JW) transformations of these terms are shown in grey. Conjugate terms are not explicitly visualised. Simulating the 20-site lattice shown requires $40$ qubits - one for each spin occupancy of each site.}
    \label{fig:FHlattice_schematic}
\end{figure*}
We have thus far shown that fast, microsecond entangling gates can be performed between ions in large 2D microtrap arrays with high-fidelity. It is thus natural to investigate the use of this platform for a large scale quantum computation. In particular, we will now describe how such a platform can be used to realise a digital simulation of a $40$-mode Fermi-Hubbard Hamiltonian, following the general approach outlined in Ref.~\cite{Lamata2014}. This is an extension of a previous analysis that investigated the use of fast gates in a $40$ ion chain in a single Paul trap to perform this task \cite{Taylor2017}; the authors found that the repetition rate requirements to perform this computation was well beyond the capabilities of current experiments. We find that this same task has a far smaller repetition rate requirement when performed on a two-dimensional microtrap array, well within the scope of current experimental demonstrations. \par
\subsection{Simulation algorithm}
The Fermi-Hubbard Hamiltonian we will seek to simulate has the following form;
\begin{equation}    \label{eqn:FH_Hamiltonian}
    H = w\sum_{\langle i,j \rangle, \sigma}^{20}(\hat{b}^\dag_{i,\sigma}\hat{b}_{j,\sigma} + \textnormal{h.c.}) + U\sum^{20}_{j=1}\hat{b}^\dag_{j,\uparrow}\hat{b}_{j,\uparrow}\hat{b}^\dag_{j,\downarrow}\hat{b}_{j,\downarrow} \,,
\end{equation}
where $\hat{b}^\dag$ $\left(\hat{b}\right)$ is the fermionic creation (annihilation) operator, $\sigma = \uparrow,\, \downarrow$ is a spin index, and $\langle i,j \rangle$ denotes nearest-neighbour pairing. This Hamiltonian describes spin-$\frac{1}{2}$ fermions in a $5\times4$ lattice, interacting with nearest-neighbours only   The first term represents tunnelling of fermions between neighbouring sites, and the second term describes the potential generated by on-site occupation. \par
We map Eq.~\eqref{eqn:FH_Hamiltonian} onto Pauli operators $\{\sigma_x,\sigma_y,\sigma_z\}$ that act on a system of $40$ interacting qubits (one for each fermionic mode) by the Jordan-Wigner (JW) transform \cite{JWtransform}:
\begin{align}
 \label{eqn:mapped_FH_ham}
    \begin{aligned}
     H \rightarrow \sum_j H_j =   w \sum_{\lambda=x,y}\bigg{(}\sum^{39}_{j=2}\sigma_{\lambda}^{j-1} \otimes  \sigma_z^j\otimes \sigma_\lambda^{j+1} \\ + \, \sum^{30}_{j=1} \sigma_{\lambda}^{j} \overset{j+9}{\underset{k=j+1}{\bigotimes}}  \sigma_z^k\otimes \sigma_\lambda^{j+10} \bigg{)} \\ 
       + \, U\left(20 + \sum^{20}_{j=1}\sigma_z^{2j}\otimes\sigma_z^{2j-1}+\sum_{k=1}^{40}\sigma_z^{k} \right) \,.
      \end{aligned}
\end{align}
Details of this mapping are included in Appendix \ref{append:JWtransform}. For purposes of digital simulation we compose the time-evolution operator by exponentiation of Eq.~\eqref{eqn:mapped_FH_ham} and employ a first-order Trotter-Suzuki decomposition (in units where $\hbar=1$):
\begin{align}
\begin{aligned}
        \hat{U}(t) = e^{-i\sum_j \hat{H}_j t} \approx \bigg{(}\prod_j e^{-i \hat{H}_j \frac{t}{n}} \bigg{)}^n \,,
    \end{aligned}
\end{align}
where $n$ is the number of Trotter steps, and the summands $H_j$ refer to individual elements of the sums in Eq.~\eqref{eqn:mapped_FH_ham}. Following the algorithm outlined in Ref.~\cite{Lamata2014}, each of the unitaries can be implemented by a local rotation on some $m$-th qubit, and a pair of multi-qubit entangling gates:
\begin{equation}
\label{eqn:unitary2umq}
     e^{-i \hat{H}_j \frac{t}{n}} = \hat{U}^\dag_\text{UMQ}\:e^{-i\phi \hat{\sigma}_z^m}\:\hat{U}_\text{UMQ} \,,
\end{equation}
where $\hat{U}_\text{UMQ}$ is the \textit{ultrafast multi-qubit} (UMQ) gate,
\begin{equation}
    \label{eqn:UMQ}
    \hat{U}_\text{UMQ} = \exp\left(-i\frac{\pi}{4}\hat{\sigma}_z^m\sum_{j \neq m} \hat{\sigma}_z^j\right) \,.
\end{equation}
The summation in the above equation is only over the qubits acted on by the particular $H_j$ term. \par
A $N$-body UMQ operation can be physically realised by a set of $N-1$ fast geometric phase (GP) gates, as well as single-qubit rotations that may be required to realise $\sigma_x$ or $\sigma_y$ couplings \cite{Lamata2014,Taylor2017}:
\begin{equation}
  \hat{U}_\text{UMQ}= e^{-i\frac{\pi}{4}\hat{\sigma}_z^m\sum_{j \neq m} \hat{\sigma}_z^j} = \prod_{j\neq m}\hat{U}_\text{GP}^{j,m}
\end{equation}
However, to construct some of the terms that arise in Eq.~\eqref{eqn:mapped_FH_ham}, non-neighbouring qubits will need to be entangled. In principle fast gates can be performed directly between non-neighbouring ions, however as the coupling of these ions to their shared motional modes decays strongly with the distance between them, it is almost always preferable to construct such operations with a series of nearest-neighbour gates \cite{Bentley2015b}. We use SWAP operations that `swap' the qubit states of two ions to connect non-neighbouring qubits; each SWAP gate can be constructed from $3$ fast gates, up to local single-qubit operations \cite{Taylor2017}. For large computations, the gates required to construct SWAP operations are likely to be the majority of the total number of entangling gates required.

\subsection{Reduction in number of gates from 1D ion chains}

As our platform for implementing this simulation algorithm, we consider a 2D microtrap array of ions as shown in Fig.~\ref{fig:FHlattice_schematic}. In this array, the numbering of qubits to ions in the lattice is chosen judiciously for specific terms that arise in Eq.~\eqref{eqn:mapped_FH_ham}. 
The aim of this section is to calculate the number of fast gates required to implement the simulation algorithm per Trotter step and compare to the number of gates required with a 1D ion chain. We will not keep track of single-qubit operations which only require one or few laser pulses to implement and can thus be performed much faster and with higher fidelity than fast two-qubit gates which typically require tens or hundreds of pulses each \cite{Taylor2017}. \par
There are three types of $H_j$ terms in the JW mapped Hamiltonian \eqref{eqn:mapped_FH_ham}: $20$ two-body terms of the form $\sigma_z^j\otimes\sigma_z^{j+1}$, $64$ three-body terms of the form $\sigma_\lambda^{j-1}\otimes\sigma_z^j\otimes\sigma_\lambda^{j+1}$, and $60$ eleven-body terms of the form  $\sigma_{\lambda}^{j} \overset{j+9}{\underset{k=j+1}{\bigotimes}}  \sigma_z^k\otimes \sigma_\lambda^{j+10}$ ($\lambda=x,y$). The unitaries arising from the two-body terms can each be implemented with a single fast gate between NN ions, not requiring the more elaborate decomposition in Eq.~\eqref{eqn:unitary2umq}.\par
The unitaries arising from three-body terms do, however, require the use of UMQ gates to implement. For example, consider the term $\sigma_x^{j-1}\otimes\sigma_z^j\otimes\sigma_x^{j+1}$, which requires the three-body UMQ gate:
\begin{equation}
\label{eqn:3bodyUMQ}
    \hat{U}_\text{UMQ}^{3\text{body}}=e^{-i\frac{\pi}{4}(\sigma_x^{j}\otimes\sigma_z^{j-1}+\sigma_z^j\otimes\sigma_x^{j+1})}\,.
\end{equation}
This operation can be realised by a pair of fast gates, $\hat{U}_\text{UMQ}^{3\text{body}}=\hat{U}_\text{GP}^{j,j-1}\hat{U}_\text{GP}^{j,j+1} $ up to single qubit rotations on $j-1$ and $j+1$ to transform from the $\sigma_z$ to the $\sigma_x$ basis. We have chosen the lattice numbering (\textit{c.f.} Fig.~\ref{fig:FHlattice_schematic}) such that the qubit pairs $(j,j-1)$ and $(j,j+1)$ correspond to nearest-neighbour ions, and thus no SWAP operations are required for the three-body UMQs. There are $64$ three-body terms in Eq.~\eqref{eqn:mapped_FH_ham}, each requiring a forward ($\hat{U}_\textnormal{UMQ}$) and backward ($\hat{U}_\textnormal{UMQ}^\dag$) UMQ to implement; in total resulting in $256$ fast gates required, per Trotter step. \par
The eleven-body terms also require UMQs, which take the form:
\begin{equation}
    \hat{U}_\text{UMQ}^{11\text{body}}=e^{-i\frac{\pi}{4}(\sigma_x^{j}\otimes\sigma_z^{j+1}+\sigma_x^{j}\otimes\sigma_z^{j+2}+\dots+\sigma_x^{j}\otimes\sigma_x^{j+10})}\,.
\end{equation}
Unlike Eq.~\eqref{eqn:3bodyUMQ}, this operation involves non-local couplings and thus its construction from fast two-qubit gates requires the use of SWAP gates to create couplings between qubits in non-neighbouring ions. The number of SWAPs required varies between the different eleven-body terms and the locations of the ions to be coupled in the array; in total the number of fast two-qubit gates required for simulation of the eleven-body Hamiltonian terms is $2352$, per Trotter step.  \par
This brings the total number of gates required for implementing the simulation on a 2D microtrap array to $2628$ gates per Trotter step (this includes 344 diagonal operations). For comparison, we now consider how many gates would be required on a 1D ion chain where qubits are numbered sequentially. The contribution from the two-body and three-body terms does not change, as they similarly can be realised without the use of SWAP operations. The UMQs that implement the eleven-body couplings will require $9$ SWAP operations, and $10$ additional two-qubit gates. In total, implementing the simulation algorithm on a 1D ion chain requires $4716$ two-qubit fast gates, per Trotter step. This demonstrates the usefulness of microtrap ion arrays for large-scale computation; the increased closeness of ions in two-dimensions allows for a $\sqrt{3}$ reduction in the total number of gates required for realising this digital simulation of the Fermi-Hubbard model. For larger-scale computations involving some $N$ qubits, this improvement will scale as $\sqrt{N}$.
\subsection{Feasibility considerations}
We now investigate the feasibility of realistically implementing this computation on a microtray array as studied in Section \ref{sec:microtrap_array}. In order for an implementation of this simulation with $10$ Trotter steps to achieve reasonable fidelity ($\gtrsim 75 \%$), individual entangling gate infidelity needs to be on the order of $10^{-5}$ or lower. As demonstrated in Fig.~\ref{fig:MicrotrapScaling} much higher fidelities $F\sim1-10^{-8}$ are possible even in large arrays. \par
For realistic trap parameters ($\omega_t=1.2$~MHz and $d=100\,\mu$m), these gates can be implemented in $1.7\mu$s with a ${\sim}500$~MHz laser. For $10$ Trotter steps, this means a total simulation time of ${
\sim}50$~ms. In contrast with the repetition rate requirements specified in Ref.~\cite{Taylor2017} (${\sim}20$~GHz), this is a far more achievable goal, and much larger repetition rate lasers have been experimentally demonstrated \cite{Heinrich2019}. This is two orders of magnitude faster than the lifetime of the metastable $D_{5/2}$ state in $ \text{Ca}^{40+}$ (considered as the qubit $\ket{1}$ state in this manuscript). Moreover, the simulation fidelity is unlikely to be affected if the rate of trap heating can be kept below $5~\text{phonons/s}$ such that phonon absorption is unlikely during the gate sequence \cite{Taylor2017}.  \par

However, the aforementioned gate fidelity of $1-10^{-8}$ assumes an idealised trap and perfect laser control. Most importantly, it does not take into account imperfect laser pulses driving $\Theta \neq \pi$ single-qubit rotations, which we have previously identified as a key experimental limitation \cite{Gale2020}. For a characteristic pulse rotation error of $\epsilon$, we have shown in Ref.~\cite{Gale2020} that a realistic estimate of single gate fidelity is:
\begin{equation}
    F = |1-N_p \epsilon|^2 \,F_0\,,
\end{equation}
where $N_p$ is the number of pulse pairs in the gate, and $F_0$ is its raw theoretical fidelity assuming perfect pulses. The gate scheme presented in Fig.~\ref{fig:MicrotrapScaling} is made up of $450$ pulse pairs, and thus rotation errors from imperfect pulses need to be on the order of $\epsilon \approx 10^{-8}$ or lower. Given that the state of the art of single-qubit gates with ultrafast pulses has a fidelity error on the order of $10^{-2}$ \cite{Campbell2010,Wong-Campos2017,Heinrich2019}, it is clear that the required experimental regime for realising such a simulation with fast gates has yet to be achieved. In Ref.~\cite{Gale2020}, we have made suggestions for improving these errors. One promising approach is to replace each pulse with a composite BB1 sequence \cite{MerrillBrown2014}, which will result in a gate scheme that is robust to first and higher order fluctuations in laser intensity.\par

\subsection{Other error sources}
While the required level of pulse control is likely the main limitation of the fast gate mechanism, experiments have a range of other limitations. We have analysed the effects of many error sources across several manuscripts; here we list some of the key points for the readers convenience. 

\textit{Stray fields.} Stray electric fields may result in differences in frequency between neighbouring traps, and is a possible source of error for fast gates in microtrap architectures. This has been studied in Ref.~\cite{Ratcliffe2020} where the authors reported that fast gate fidelity is robust to stray fields as large as ${\sim}1\%$ of the applied voltages. 

\textit{Hot motional states.} In the same manuscript, the authors demonstrated that these fast gate schemes are robust to high temperatures, even for mean motional occupations on the order of $\Bar{n}=10-100$ \cite{Ratcliffe2020}.

\textit{Trap heating.} Ref.~\cite{Taylor2017} investigated the effect of trap heating, discovering that gate fidelity is only significantly affected if a heating event occurs during the gate operation. Effectively, the heating rate places a limit on the total number of gates that can be sequentially implemented. However, given the gate speeds reported in this manuscript, this number may still be large enough for large-scale computation for experimentally demonstrated heating rates (as discussed earlier in this section). 

\textit{Timing errors.} The effect of errors on the pulse timings have been discussed previously, both for Gaussian noise \cite{Ratcliffe2018a} and shot-to-shot fluctuations Ref.~\cite{Mehdi2020Chain}. State-of-the-art lasers have exceptionally stable repetition rates, with Refs.~\cite{Hussain2016,Heinrich2019} demonstrating fractional instabilities below $10^{-7}$. The gates discussed in this manuscript can be made robust to timing errors, by fixing pulse timings to integer multiples of the repetition period. This can be done by adding a second stage to the optimisation protocol wherein the pulse timings are optimised on a finite grid specified by the repetition rate, as we have demonstrated in Refs~\cite{Gale2020,Mehdi2020Chain}.

\section{Acknowledgments}
This research was undertaken with the assistance of resources and services from the National Computational Infrastructure (NCI), which is supported by the Australian Government.

\section{Appendices}
\appendix

\section{Non-dimensionalisation mode structure \label{appendix:non-dimensionalisation}}

\begin{figure*}
    \centering
    \includegraphics[width=0.8\textwidth]{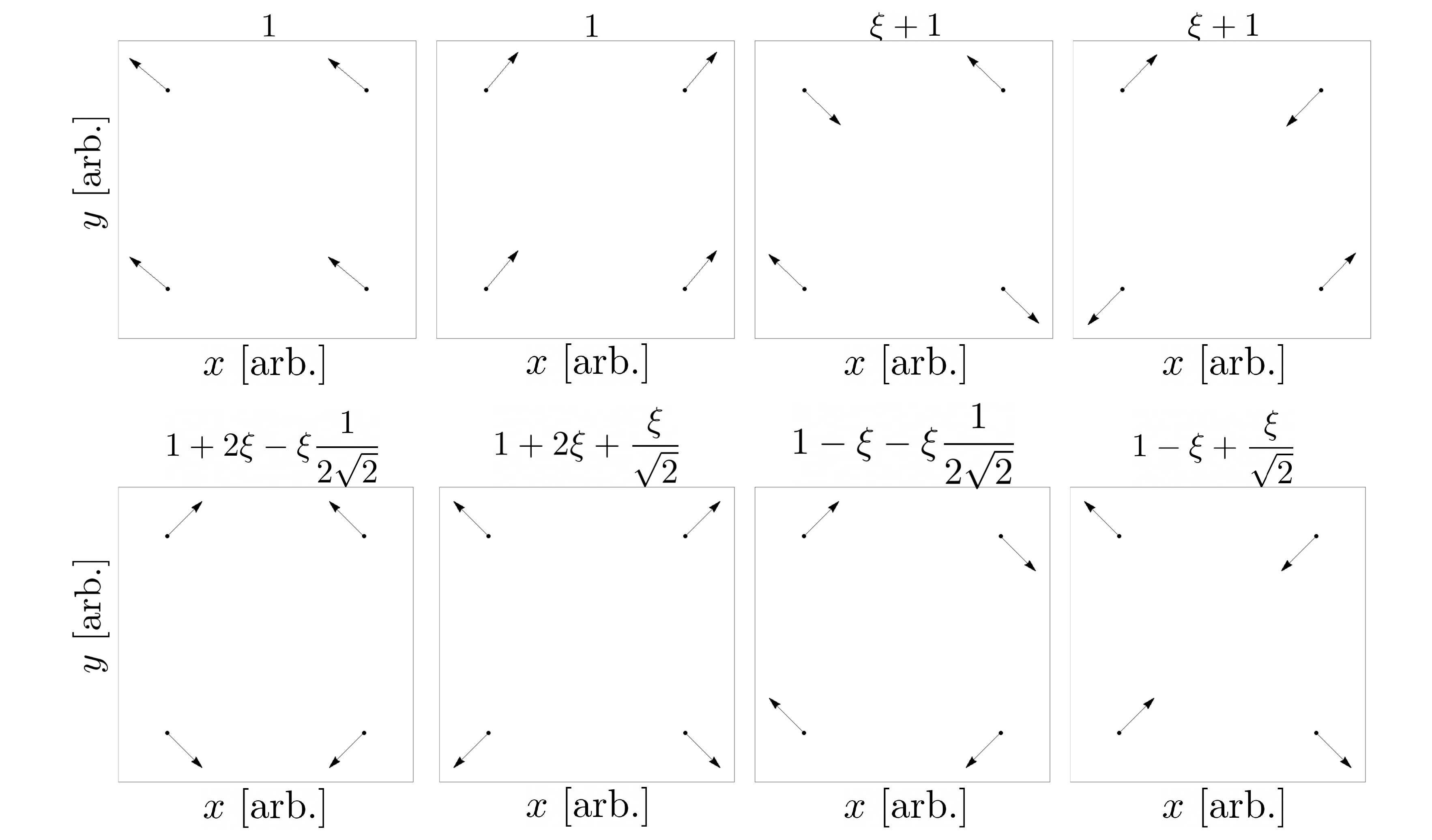}
    \caption{Visualisation of the couplings of the motional modes to the ions in a $2\times 2$ square cell. The displacements in the $x$ and $y$ directions are represented in arbitrary units as the coupling vectors $\mathbf{b_m}$ are normalised. The corresponding squared frequency of each mode is annotated, $\omega_m^2/\omega_t^2$.}
    \label{fig:2D_sym_modes}
\end{figure*}

Visualisations of the corresponding mode eigenvectors, $\mathbf{b_m}$, are shown in Figure \ref{fig:2D_sym_modes}.
Due to the symmetries of the microtrap array we have considered, we are able to find non-dimensional expressions for the mode frequencies of a square $2\times2$ cell of ions in terms of the non-dimensional parameter $\xi\equiv\frac{\omega_{BR}^2-\omega_t^2}{\omega_t^2}$. There are $8$ relevant motional modes of this system, with frequencies $\omega_m$ that can be expressed as:
\begin{align}
\label{eqn:2D_modefreq_analytical}
\begin{aligned}
    \frac{\omega_m^2}{\omega_t^2} = \bigg{\{}1,1,\xi +1,\xi +1,-\frac{\xi }{2 \sqrt{2}}-\xi +1,-\frac{\xi }{2 \sqrt{2}}+2 \xi +1, \\
    \frac{\xi }{\sqrt{2}}-\xi +1,\frac{\xi }{\sqrt{2}}+2 \xi +1  \bigg{\}} \,.
\end{aligned}
\end{align}
This parameter can be expressed in terms of the inter-trap distance $d$ and trap frequency $\omega_t$:
\begin{align}
\begin{aligned}
    \xi =\frac{2 \Lambda }{27}\left(\frac{\sqrt[3]{\delta }+\frac{18-8 \sqrt{2}}{\sqrt[3]{\delta }}+2 \left(\sqrt{2}-4\right)}{3 \sqrt{2} \left(2 \sqrt{2}-1\right)}+1\right)^{-3} \,,
\end{aligned}
\end{align}
where
\begin{align}
\begin{aligned}
\delta =\sqrt{-3528 \sqrt{2} \Lambda ^2+5537 \Lambda ^2-11228 \sqrt{2} \Lambda +16688 \Lambda }\\
-28 \sqrt{2} \Lambda +63 \Lambda -50 \sqrt{2}+88 \,,
\end{aligned}
\end{align}
\begin{equation}
    \Lambda =\frac{1}{M}\frac{ 27 }{d^3 \omega_t ^2}\frac{e^2}{4\pi\epsilon_0} \,.
\end{equation}
The components corresponding mode eigenvectors $\mathbf{b}_m$ can be interpreted as the couplings of the ions positions to the $m$-th motional mode. These vectors are visualised for each of the modes of the $2\times 2$ microtrap array in Fig.~\ref{fig:2D_sym_modes}.

\section{\label{append:optmethods}Optimisation methods}
Our approach to gate design employs numerical optimisation techniques to identify a pulse sequence that implements the desired entangling operation with fidelity as close to unit value as possible. Specifically, we utilise the global optimisation methods outlined in Ref.~\cite{Gale2020} to numerically minimize an expression of gate infidelity. We use a truncated expression (generalised from Refs.~\cite{Ratcliffe2018a,Ratcliffe2020,Gale2020}) for the infidelity of a fast entangling gate between two ions, $\mu$ and $\nu$, 
\begin{align}
    \label{eqn:2d infidelity truncated}
     1-F \approx \frac{2}{3}|\Delta \phi| ^2 + \frac{4}{3}\sum_m (\frac{1}{2}+\Bar{n}_m) \big( (\mathbf{K_\mu} \cdot \mathbf{b_m})^2 \\ \nonumber + (\mathbf{K_\nu}\cdot\mathbf{b_m})^2 \big)|\Delta \alpha_m|^2 \,,
\end{align}
where $\Bar{n}_m$ and $\mathbf{b_m}$ are, respectively, the average phonon occupation and classical eigenvector of the $m$-th motional mode,  $\Delta\phi$ is the phase-mismatch, and $\Delta \alpha_m$ is the unrestored motion of the $m$-th motional mode in phase-space:
\begin{align}
    \label{eqn:dphi expression}
    \Delta \phi = \bigg{|}8\eta_m^2 (\mathbf{K_\mu} \cdot \mathbf{b_m})(\mathbf{K_\nu} \cdot \mathbf{b_m})\sum_{i\neq j} z_i z_j \\\nonumber \sin{(\omega_m |t_i-t_j|)}\bigg{|} -\frac{\pi}{4} \,,
\end{align}
\begin{equation}
    \label{eqn:dPp expression}
    \Delta \alpha_m = 2\eta_m \sum_{k = 1} z_k e^{-i \omega_m t_k} \,.
\end{equation}
where $z_k$ and $t_k$ are the number of pulse pairs and the time of arrival at the ion of the $k$-th pulse group, respectively. Here $\eta_m=k\sqrt{\frac{\hbar}{2M\omega_m}}$ is the Lamb-Dicke parameter of the $m$-th motional mode and $\mathbf{K_i}$ is a normalised vector corresponding to the direction of the laser pulses with respect to the coordinates of the $i$-th ion. The inclusion of this vector allows the coupling of the laser-light to the motional modes to be expressed for arbitrary laser orientation in the $x-y$ plane. We assume a coordinate basis where the $x,y$ co-oordinates correspond to the row and columns of the ion array, i.e.
\begin{equation}
    \label{ion position variables}
    \mathbf{x}_{i,j}=\{x_{1,1},y_{1,1},x_{1,2},y_{1,2},x_{2,1},y_{2,1},x_{2,2},y_{2,2}\},
\end{equation}
where $(i,j)$ corresponds to the $i$-th ion along the $j$-th row of the array. Unless otherwise stated, we will assume the direction of the laser pulses to be co-linear with the line that passes through the equilibrium positions of the two ions involved in the gate; i.e. the two ions are kicked directly towards/away from each other by each kick from the counterpropagating pulse pair. It is worth noting that the next order terms of the infidelity are strictly negative, and thus Eq.~\eqref{eqn:2d infidelity truncated} provides a lower-bound on the theoretical fidelity of a particular pulse sequence. \par

For efficient optimisation in the presence of multiple surrounding ions, we optimise an anti-symmetric pulse sequence with $16$ pulse groups that arrive at the ions at regular intervals, i.e:
\begin{align}
\label{eq:APG(16)_zandt}
\begin{aligned}
    \mathbf{z}_k &= \left\{ -z_{8}, \, \dots, \, -z_2, \, -z_1, \, z_1, \, z_2, \, \dots, \, z_{8} \right\} \, , \\
    \mathbf{t}_k &= \frac{T_G}{16} \left\{ -8, \, \dots, \, -2, \, -1, \, 1, \, 2, \, \dots, \, 8 \right\} \, .
\end{aligned}
\end{align}
where $T_G$ is the total gate operation time. This is known as the APG($16$) scheme, where numerical optimisation is performed over the elements $\{z_1,\,\dots,z_8\}$. The anti-symmetric constraints on the elements of $\mathbf{z}$ and $\mathbf{t}$ guarantee momentum restoration of each motional mode, reducing the expression of $|\Delta \alpha_m|$ to
\begin{equation}
    |\Delta \alpha_m| = 2\eta_m  \sum_{k = 1}^{N} z_k \sin(\omega_m t_k) \,.
\end{equation}
This reduces the complexity of the infidelity cost-function, Eq.~\eqref{eqn:2d infidelity truncated}, which is desirable for optimisations in 2D architectures where even the most simple arrays have many motional modes. For further details on the numerical optimisation routine used, see Ref.~\cite{Gale2020}. \par

Here we provide two examples of optimised gate sequences, the fidelity of which can be verified using the expressions above. The first is for a system characterised by $\xi=1.2\times10^{-4}$:
\begin{align*}
    \mathbf{z}_k &= \{23,-47,-47,-32,31,-41,-47,-38,\\
    & 38,47,41,-31,32,47,47,-23\} \, , \\
    \mathbf{t}_k &= \{-1,-\frac{7}{8},-\frac{3}{4},-\frac{5}{8},-\frac{1}{2},-\frac{3}{8},-\frac{1}{4},-\frac{1}{8}, \\
    & \frac{1}{8},\frac{1}{4},\frac{3}{8},\frac{1}{2},\frac{5}{8},\frac{3}{4},\frac{7}{8},1\}\tau_0 \, ,
\end{align*}
which describes the gate with $1-F\simeq 10^{-9}$, $f_\text{min}=450\frac{\omega_t}{2\pi}$, and $T_G=2.0\tau_0$. This is the gate used to create Figures \ref{fig:PS_traj} and \ref{fig:MicrotrapScaling}. The second example is, again for a system characterised by $\xi=1.2\times10^{-4}$:
\begin{align*}
    \mathbf{z}_k &= \{-74,-41,-14,66,24,-30,-72,-76,\\
   & 76,72,30,-24,-66,14,41,74\} \, , \\
    \mathbf{t}_k &= \frac{1}{16}\{-10,-\frac{35}{4},-\frac{15}{2},-\frac{25}{4},-5,-\frac{15}{4},-\frac{5}{2},-\frac{5}{4},\\
    & \frac{5}{4},\frac{5}{2},\frac{15}{4},5,\frac{25}{4},\frac{15}{2},\frac{35}{4},10\}\tau_0 \, .
\end{align*}
This describes a gate with $1-F\simeq 10^{-4}$, $f_\text{min}=950\frac{\omega_t}{2\pi}$, and $T_G=1.25\tau_0$.

During this optimisation procedure, and in the fidelities we report, we assume the Coulomb interaction can be truncated to second order in the ion co-ordinates. This is an assumption that will generally impact the gate fidelity, but can be corrected for in a second-stage of optimisation which uses an ODE description of the ions motional dynamics \cite{Gale2020}. This ODE description is also able to explicitly incorporate the finite laser repetition rate. As this optimisation is a simple extension of the two-ion example presented in Ref.~\cite{Gale2020}, we have not explicitly included these corrections in this manuscript.

\begin{widetext}
\section{\label{append:JWtransform} Jordan-Wigner transformation of Fermi-Hubbard Hamiltonian}
In this Appendix, we calculate the Jordan-Wigner (JW) transformation of the fermionic Hamiltonian in Eq.~\eqref{eqn:FH_Hamiltonian}. The Jordan-Wigner transformation can be expressed in the form
\begin{eqnarray}
    \label{JW transform}
    b_j &=& -\left(\overset{j-1}{\underset{n=1}{\bigotimes}} \, \sigma_z^n \right) \otimes \sigma_{-}^j \\
    b_j^\dag &=& -\left(\overset{j-1}{\underset{n=1}{\bigotimes}}\, \sigma_z^n \right) \otimes \sigma_{+}^j \,,
\end{eqnarray}
and it can be simply verified that the transformation maintains the anti-commutation relations $\{b_j,b_k^\dag\}=\delta_{jk}$. The fermionic operators in Eq.~\eqref{eqn:FH_Hamiltonian} need to first be indexed by only a single value, and thus we will re-index $ b_{j,\uparrow}\rightarrow b_{2j},\, b_{j,\downarrow} \rightarrow b_{2j-1}$. \par
We begin by considering the on-site interaction terms that are of the form $U\hat{b}^\dag_{j,\uparrow}\hat{b}_{j,\uparrow}\hat{b}^\dag_{j,\downarrow}\hat{b}_{j,\downarrow}$. As these terms only contain pairs of creation/annihilation operators, the $\sigma_z$ part of the transformation will cancel as $(\sigma_z)^2 = \mathbf{1}$:
\begin{eqnarray}
    U\sum^{20}_{j=1}b^\dag_{j,\uparrow}b_{j,\uparrow}b^\dag_{j,\downarrow}b_{j,\downarrow} &\rightarrow& U\sum^{20}_{j=1}(\sigma_+^{2j}\otimes\sigma_-^{2j})\otimes(\sigma_+^{2j-1}\otimes\sigma_-^{2j-1}) \\
    &=& \frac{U}{4} \sum^{20}_{j=1}\left(1+\sigma_z^{2j} \right)\otimes\left(1+\sigma_z^{2j-1} \right)
    \\ 
    &=& \frac{U}{4}\left(20 + \sum^{20}_{j=1}\sigma_z^{2j}\otimes\sigma_z^{2j-1}+\sum_{k=1}^{40}\sigma_z^{k} \right) \,,
    \label{JW_onsite}
\end{eqnarray}
where we have used $\sigma_+\otimes\sigma_- = \frac{1}{2}(1+\sigma_z)$ in the second line. \par
The next terms we will consider are those that describe tunnelling of spins between neighbouring sites along a row of the lattice. We will consider the example of the term $w\,b^\dag_{j,\uparrow}b_{j-1,\uparrow}+\textnormal{h.c.}$:
\begin{equation}
    w b^\dag_{j,\uparrow}b_{j-1,\uparrow}+\textnormal{h.c.} \rightarrow w\,b^\dag_{2j}b_{2j-2}+\textnormal{h.c.} \,.
\end{equation}
By using the identity $(\sigma_{z})^{2} = \mathbf{1}$, the JW mapping of this can be expressed as:
\begin{eqnarray}
   w b^\dag_{2j,\uparrow}b_{2j-2,\uparrow}+\textnormal{h.c.} &\rightarrow& w\left(\sigma_-^{2j-2} \otimes \sigma_z^{2j-1}\otimes\sigma_+^{2j}  \right) + \text{h.c.} \\
    &=& \frac{w}{2}\left(\sigma_{x}^{2j-2} \otimes  \sigma_z^{2j-1}\otimes \sigma_x^{2j}+\sigma_{y}^{2j-2} \otimes  \sigma_z^{2j-1}\otimes \sigma_y^{2j} \right)\,,
    \label{JW_row}
\end{eqnarray}    
where in the last line we have expanded $\sigma_\pm = \frac{1}{2}(\sigma_x\pm i \sigma_y)$. \par
Finally, we have terms such as $w\,b^\dag_{6,\uparrow}b_{1,\uparrow}+\textnormal{h.c.}$ which correspond to tunnelling of spins between nearest-neighbour sites along columns of the Fermi-Hubbard lattice. This mapping is similar to the row-tunnelling term considered above, with fewer cancellations arising from $(\sigma_{z})^{2} = \mathbf{1}$. For example, for the term $b^\dag_{1,\uparrow}b_{6,\uparrow} + \textnormal{h.c.}$, the JW mapping is
\begin{eqnarray}
    b^\dag_{1,\uparrow}b_{6,\uparrow} + \textnormal{h.c.} &\rightarrow& b^\dag_{2}b_{12} + \textnormal{h.c.} \\ &=& \sigma_{-}^2 \otimes  \sigma_z^3\otimes \dots \otimes \sigma_z^{11} \otimes \sigma_+^{12} + \textnormal{h.c.} \\ \nonumber
    &=& \frac{1}{2}\left( \sigma_{x}^2 \otimes  \sigma_z^3\otimes \dots \otimes \sigma_z^{11} \otimes \sigma_x^{12} + \sigma_{y}^2 \otimes  \sigma_z^3\otimes \dots \otimes \sigma_z^{11} \otimes \sigma_y^{12} \right) \,.
\end{eqnarray}
For a general column-tunnelling term $b^\dag_{j,\uparrow}b_{j+5,\uparrow} + \textnormal{h.c.}$ the JW mapping has the form:
\begin{equation}
     b^\dag_{j,\uparrow}b_{j+5,\uparrow} + \textnormal{h.c.} \rightarrow \frac{w}{2} \left(\sigma_{x}^{j} \overset{j+9}{\underset{k=j+1}{\bigotimes}}  \sigma_z^k\otimes \sigma_x^{j+10}+\sigma_{y}^{j} \overset{j+9}{\underset{k=j+1}{\bigotimes}}  \sigma_z^k\otimes \sigma_y^{j+10} \right) \,.
     \label{JW_col}
\end{equation}
Combining the mappings expressed in Equations \eqref{JW_onsite}, \eqref{JW_row}, and \eqref{JW_col}, we can express the full JW mapping of the Fermi-Hubbard Hamiltonian \eqref{eqn:FH_Hamiltonian}:
\begin{eqnarray}
    H &=& w \sum_{\lambda=x,y}\left(\sum^{39}_{j=2}\sigma_{\lambda}^{j-1} \otimes  \sigma_z^j\otimes \sigma_\lambda^{j+1} + \sum^{30}_{j=1} \sigma_{\lambda}^{j} \overset{j+9}{\underset{k=j+1}{\bigotimes}}  \sigma_z^k\otimes \sigma_\lambda^{j+10} \right) +U\left(20 + \sum^{20}_{j}\sigma_z^{2j}\otimes\sigma_z^{2j-1}+\sum_{k}^{40}\sigma_z^{k} \right) \,,
\end{eqnarray}
where we have re-scaled $U{\rightarrow}4U$ and $w{\rightarrow}2w$ for convenience.

\end{widetext}

\bibliographystyle{bibsty}
\bibliography{bibliography}

\end{document}